\documentclass[aps,prx,reprint,superscriptaddress, twocolumns]{revtex4-2}
\usepackage{amsmath,amssymb}
\usepackage{bbold}
\usepackage{float}
\usepackage{graphicx}
\usepackage{dcolumn}
\usepackage{bm}
\usepackage[colorlinks]{hyperref}
\usepackage{braket}
\usepackage{mathtools}
\usepackage{comment}
\usepackage{xcolor}
\usepackage{physics}
\usepackage{enumerate}

\colorlet{alena}{blue!80!gray}
\colorlet{roman}{green!40!gray}

\begin{document}

\title{Tensor Quantum Programming}

\author{A. Termanova}
\altaffiliation{These authors contributed equally: A.~Termanova and Ar.~Melnikov.}
\affiliation{Terra Quantum AG, Kornhausstrasse 25, 9000 St. Gallen, Switzerland}

\author{Ar. Melnikov} 
\altaffiliation{These authors contributed equally: A.~Termanova and Ar.~Melnikov.}
\affiliation{Terra Quantum AG, Kornhausstrasse 25, 9000 St. Gallen, Switzerland}

\author{E. Mamenchikov}
\affiliation{Terra Quantum AG, Kornhausstrasse 25, 9000 St. Gallen, Switzerland}

\author{N. Belokonev}
\affiliation{Terra Quantum AG, Kornhausstrasse 25, 9000 St. Gallen, Switzerland}

\author{S. Dolgov}
\affiliation{Terra Quantum AG, Kornhausstrasse 25, 9000 St. Gallen, Switzerland}

\author{A. Berezutskii}
\affiliation{Terra Quantum AG, Kornhausstrasse 25, 9000 St. Gallen, Switzerland}

\author{R. Ellerbrock}
\affiliation{Terra Quantum AG, Kornhausstrasse 25, 9000 St. Gallen, Switzerland}

\author{C. Mansell}
\affiliation{Terra Quantum AG, Kornhausstrasse 25, 9000 St. Gallen, Switzerland}

\author{M. Perelshtein}
\affiliation{Terra Quantum AG, Kornhausstrasse 25, 9000 St. Gallen, Switzerland}

\begin{abstract}
Running quantum algorithms often involves implementing complex quantum circuits with such a large number of multi-qubit gates that the challenge of tackling practical applications appears daunting. 
To date, no experiments have successfully demonstrated a quantum advantage due to the ease with which the results can be adequately replicated on classical computers through the use of tensor network algorithms.
Additionally, it remains unclear even in theory where exactly these advantages are rooted within quantum systems because the logarithmic complexity commonly associated with quantum algorithms is also present in algorithms based on tensor networks.
In this article, we propose a novel approach called Tensor Quantum Programming, which leverages tensor networks for hybrid quantum computing. 
Our key insight is that the primary challenge of algorithms based on tensor networks lies in their high ranks (bond dimensions).
Quantum computing offers a potential solution to this challenge, as an ideal quantum computer can represent tensors with arbitrarily high ranks in contrast to classical counterparts, which indicates the way towards quantum advantage.
While tensor-based vector-encoding and state-readout are known procedures, the matrix-encoding required for performing matrix-vector multiplications directly on quantum devices remained unsolved.
Here, we developed an algorithm that encodes Matrix Product Operators into quantum circuits with a depth that depends linearly on the number of qubits.
It demonstrates effectiveness on up to 50 qubits for several matrices frequently encountered in differential equations, optimization problems, and quantum chemistry.
We view this work as an initial stride towards the creation of genuinely practical quantum algorithms.

\end{abstract}
\maketitle

\section{Introduction}

Quantum computing has garnered significant attention in both academia and industry. Of particular interest is hybrid quantum computing, where the main idea is to efficiently combine classical and quantum computation~\cite{endo2021hybrid, white_paper_tq}. Considerable research efforts are being devoted to areas ranging from variational algorithms~\cite{variational_algorithms} to quantum machine learning~\cite{biamonte2017quantum_ML, Stoudenmire-Quantum-ML-2019}. On the one hand, a clear understanding of how these methods provide advantages over purely classical ones is still somewhat lacking~\cite{aaronson2015read, tang2022dequantizing}. On the other hand, important insights are being made~\cite{martyn2021grand}. Despite the ongoing theoretical progress, it is still very difficult to devise algorithms that would allow even an ideal quantum computer (QC) to efficiently solve practical tasks, such as simulating physical or chemical systems~\cite{perelshtein2020large, rolls_roys}. Hence, the purpose of this work is to develop a methodology for creating effective quantum algorithms.

In the circuit model of quantum computation, quantum algorithms consist of three main steps: data loading/encoding (quantum state preparation), data processing (multi-qubit quantum gates), and measurement (quantum state tomography). The primary challenge facing gate-based quantum computing is the significant complexity involved in implementing these steps for arbitrary states and multi-qubit gates. We elaborate on these complexities later in the paper but in short, the complexity of decomposing arbitrary multi-qubit gates into single- and two-qubit gates is exponential with respect to the number of qubits~\cite{TheorLimQGdec, k-qubits_gate, QuantumShannonDecomposition, Rakyta2022approaching}, and similar results hold for the preparation and readout of arbitrary states~\cite{plesch2011_exp_prep_complex, paris2004_exp_tomography}.

Consequently, in our present work, we suggest restricting the considered states and matrices to those in the tensor train (TT) format~\cite{oseledets2011tensor}, a simple type of tensor network (TN)~\cite{tensor_networks_review}. The main reason for this is that such states and matrices possess small entanglement and therefore can be efficiently realized via shallow circuits. For example, there are protocols for implementing the encoding of a Matrix Product State (MPS) into shallow circuits~\cite{MPS_preparation, two_qubits_MPS_encoding, auto_dif_MPS_preparation, rudolph2022mps_decomposition, melnikov2023quantum}, which also work well with poor connectivity between qubits. MPS and TT-vector are identical from the mathematical point of view -- we use them interchangeably hereinafter. Moreover, any quantum circuit can be naturally regarded as a TN~\cite{markov2008simulating}. This insight leads us to shift our focus towards the naturally arising inverse task: encoding a TN into a quantum circuit. By focusing on encoding TTs, we leverage the structured nature of TNs, bypassing the complexity associated with arbitrary encoding.

The rest of the paper is organized as follows. In Sec.~\ref{sec:Preliminaries}, we present a concise exposition of the mathematical principles underlying TTs. In Sec.~\ref{sec:TQPP}, we introduce the hybrid computing concept of Tensor Quantum Programming. In Sec.~\ref{sec:MPS encoding}, we briefly review the existing algorithms for encoding MPSs into a quantum computer. In Sec.~\ref{sec:Method}, we propose a method for encoding TT-matrices into quantum circuits, which underlies our scheme of Tensor Quantum Programming. In Sec.~\ref{sec:results}, we present numerical results investigating the efficacy of our encoding algorithm. In Sec.~\ref{sec:TTDE}, we explain the strategy for achieving efficient data-readout through MPS tomography. In Sec.~\ref{sec:Applications}, we provide a detailed overview of the application domains wherein the proposed scheme of hybrid computations can offer advantages through the use of a QC. 

\section{Preliminaries}\label{sec:Preliminaries}

\begin{figure}[ht]
    \centering
    \includegraphics[width = 0.99\linewidth]{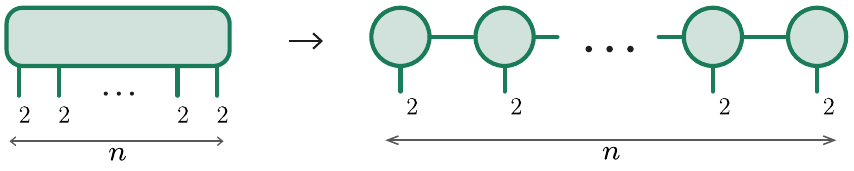}
    \caption{Graphical notation showing an $n$-dimensional tensor in its TT format. Circles and blocks indicate tensors. The dangling vertical bonds represent physical indices. Unless stated otherwise, we assume that the dimension of physical indices equals 2. The connected lines denote tensor contractions over virtual indices.}
    \label{fig:tensor_to_tt}
\end{figure}

In this section, we give a brief introduction to the mathematics behind TTs. Any $n$-dimensional tensor can be represented or approximated in TT format~\cite{oseledets2011tensor}, which is a contracted chain of $n$ lower-order tensors shown in  Fig.~\ref{fig:tensor_to_tt}. 

\begin{figure}[ht]
    \centering    \includegraphics[width = 0.8\linewidth]{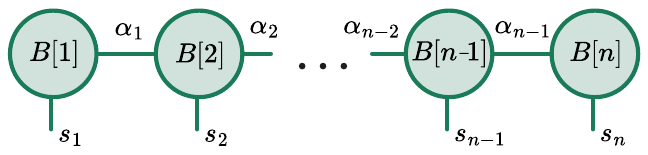}
    \caption{
    MPS representation of the state vector $\ket{\psi}$ defined in Eq.~\eqref{eq:psi_mps}.}
    \label{fig:mps}
\end{figure}

Such a decomposition is also known as the MPS format~\cite{tensor_networks_review}. The term ``state” refers to the original use of MPS to describe quantum states. One can write a quantum state $\ket{\psi}$ of $n$ qubits in the MPS format by expanding the state vector in the computational basis into a TT:
\begin{equation}\label{eq:psi_mps}
    \ket{\psi}_\text{\tiny{MPS}} = \sum\limits_{\{ s \}} B\text{\small{[1]}}^{ s_1}_{1 \alpha_1} B\text{\small{[2]}}^{ s_2}_{\alpha_1 \alpha_2} \dots B\text{\small{[n]}} ^{s_n}_{\alpha_{n - 1 }1} \ket{s_1 s_2 \dots s_n},
\end{equation}
where the terms $B\text{\small{[i]}}^{ s_i}_{\alpha_{i-1}, \alpha_i }$ are $n$ different $3$-dimensional tensors (TT-cores), and we use the Einstein summation convention that repeated indices are summed over. Each tensor contains a physical index $s_i \in \{1,2\} $ and bond indices $ \alpha_i \in \{1,2, \dots, r_i\}$, as depicted in Fig. \ref{fig:mps}. We call the maximum value of the bond index, $r$, the bond dimension, or the MPS (TT-) rank. 

A crucial connection exists between the entanglement of a quantum state and its tensor train rank. Specifically, for an MPS $\ket{\psi}$ with a rank $r$, the entropy of the reduced subsystem $\rho = \text{Tr}_2{[\ket{\psi} \bra{\psi}]}$ follows an upper bound given by $S(\rho) \leq \log{r}$~\cite{tensor_networks_review}. Therefore, MPS states are often called weakly entangled.

\begin{figure}[ht]
    \centering
    \includegraphics[width = 0.8\linewidth]{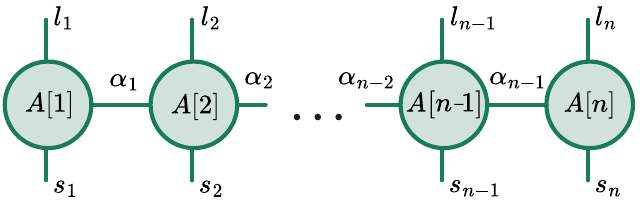}
    \caption{
    MPO representation of the linear operator $H$ defined in Eq.~\eqref{eq:H_mpo}.
    }
    \label{fig:mpo}
\end{figure}

Similar to the MPS format for vectors, the MPO (Matrix Product Operator) is a TT representation for matrices or linear operators. The MPO representation of some linear operator, $H$, on $n$-qubits reads as follows: 
\begin{multline}\label{eq:H_mpo}
        H_\text{\tiny{MPO}} = \sum\limits_{\{ s \}, \{l \}} A\text{\small{[1]}}^{ s_1 l_1}_{1 \alpha_1} A\text{\small{[2]}}^{ s_2 l_2}_{\alpha_1 \alpha_2} \dots A\text{\small{[n]}}^{ s_n l_n}_{\alpha_{n - 1} 1} \times \\ \ket{s_1 s_2 \dots s_n} \bra{l_1 l_2 \dots l_n},
\end{multline}
where now the TT-cores $A\text{\small{[i]}}^{ s_i l_i}_{\alpha_{i-1} \alpha_i}$ are $4$-index tensors with two physical indeces $s_i$ and $l_i \in \{1,2\}$, as depicted in Fig. \ref{fig:mpo}. The rank $r$ of an MPO is again the maximum value among all bond indexes $\alpha_i$.

The TT format supports algebraic operations on tensors, allowing norms, dot products, element-wise sums, and products of tensors to be computed without involving the whole tensor but working only with TT-cores~\cite{oseledets2011tensor}. This provides a computationally efficient tool for working with vectors, matrices, quantum states, and even data in machine learning~\cite{ahmadi2022cross_video_image}. For instance, discretized smooth functions are usually well-represented in the MPS format~\cite{holmes2020function_preparation}. Moreover, there is an exact MPS decomposition for trigonometric, exponential, polynomial, and rational functions~\cite{QTT_functions}. When discussing matrices, the MPO format frequently offers accurate approximations for structured matrices. Take, for instance, Toeplitz matrices, which consistently exhibit an MPO rank of $3$~\cite{kazeev2012_toeplitz}. Transforming any MPS into a diagonal MPO matrix preserves its rank~\cite{oseledets2011tensor}.

\section{Tensor Quantum Programming Paradigm}\label{sec:TQPP}

\begin{figure*}
    \centering
    \includegraphics[width = 0.9\linewidth]{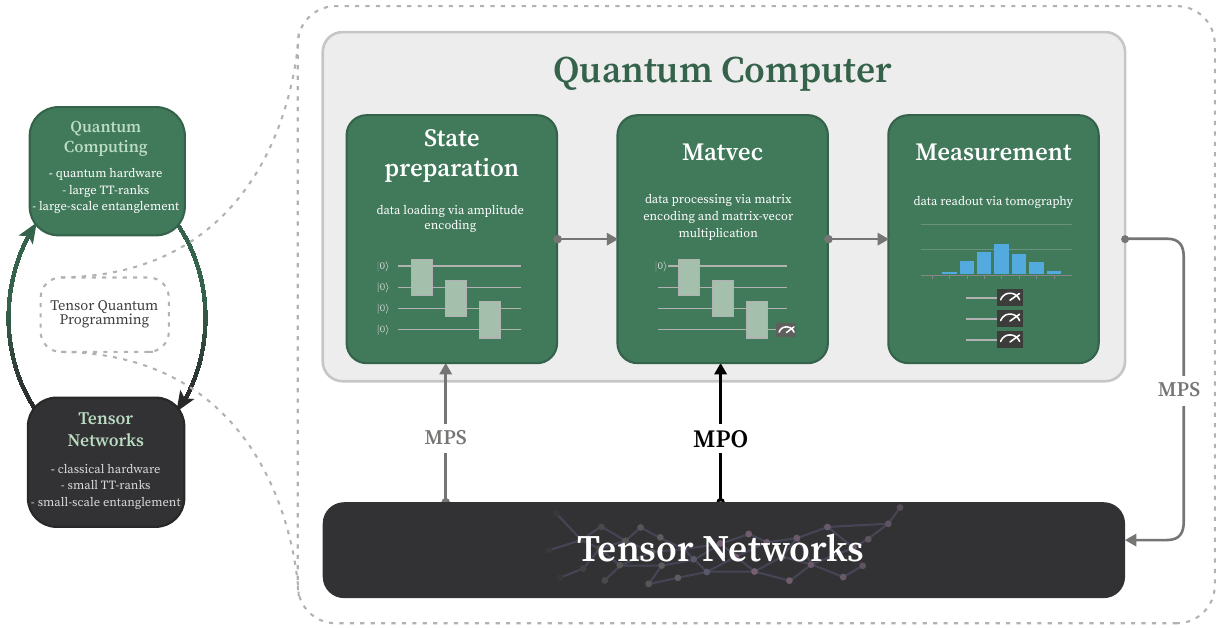}
    \caption{Our hybrid paradigm of Tensor Quantum Programming $TN \xrightarrow{} QC \xrightarrow{} TN$. 
    Quantum state preparation is performed using an MPS-encoding protocol~\cite{QPrep, melnikov2023quantum}. Matrix-vector multiplication (the main operation on a QC) is implemented with the MPO-encoding scheme presented in this work. Quantum state tomography is achieved using an MPS tomography algorithm~\cite{pastaq}. In the presented paradigm, the states at the input and output of a quantum computer are effectively represented in the form of TNs, and all of the steps have a logarithmic complexity in terms of the size of the vectors.}
    \label{fig:scheme_hybrid}
\end{figure*}

We start with the key principles of tensor networks and their connection to quantum computing. A key characteristic of MPS (TT-vector) and MPO (TT-matrix) is the rank, also known as the bond dimension, which determines the degree of entanglement or correlations in the vector or matrix. If the rank is small, then such a representation is very efficient and allows the number of parameters required for data-encoding to be compressed from exponential scaling to polynomial scaling. In addition, as mentioned in the previous section, it is possible to prepare (step 1) and read out (step 3) a MPS on a quantum computer efficiently with linear complexity in the number of qubits~\cite{melnikov2023quantum, cramer2010_tomograthy}. Remarkably, as we demonstrate below, it is possible to efficiently implement one of the main operations -- the {\tt Matvec} operation (matrix-vector multiplication, step 2) -- on a quantum computer.

Consequently, this allows us to introduce the hybrid computing concept of Tensor Quantum Programming: $TN \xrightarrow{} QC \xrightarrow{} TN$, shown in Fig.~\ref{fig:scheme_hybrid}. Firstly, we assume that an algorithm for solving a particular numerical problem is written solely in the language of tensor networks. That is, only operations with tensor networks must be present. It is worth noting that often the algorithm consists of multiple repeated iterations.  Subsequently, the classical implementation of this algorithm is initiated. This approach is chosen because, in many instances, the initial ranks tend to be small. Commencing the implementation classically proved to be considerably more efficient in these cases compared to an immediate quantum implementation. Moreover, should we observe an increase in the ranks of the solution, we proceed to transfer the vector from the intermediate solution, obtained during an intermediate iteration, to a quantum computer, employing the MPS preparation algorithm (step 1 in Fig.~\ref{fig:scheme_hybrid})~\cite{melnikov2023quantum}. The following iterations of the algorithm in the presence of large ranks of the solution are carried out on a quantum computer via the method developed in Sec.~\ref{sec: MPO_encoding}. As the ranks diminish once more, we can transfer the intermediate solution back to a classical computer and resume the process there, employing the MPS tomography algorithm~\cite{pastaq}.

This scheme enables us to identify the advantage of a quantum computer over a classical one. It is that much larger values of TT-ranks (the entanglement of the state) are accessible. That is, problems with stronger correlations can be solved since the complexity of quantum calculations is determined only by the complexity of the quantum circuit and does not depend on the state vector’s entanglement, which can grow unlimitedly during execution. 

Further, in Sec.~\ref{sec: MPO_encoding}, we show that the complexity of the circuit implementing the multiplication of the matrix encoded into it is determined by the matrix rank, which in many practical problems is often small~\cite{kornev2023_NS,  udell2019big, beckermann2019bounds, chen2023quantum}. Classically, tensor computing is limited by the ranks of all intermediate computation results, while quantum computing requires low ranks to prepare a quantum state efficiently, but while computing, the ranks can increase without bounds. Yet, for effective quantum state reading, we must reduce the ranks.
Formulated exclusively in terms of tensor networks, we opt for the most suitable hardware for each phase, depending on the size of the vector's rank: classical when the rank is small and quantum when it is large. The benefits of this overarching paradigm are discussed in Sec.~\ref{sec:conclusion}.

\section{Vector-encoding via Matrix Product States}\label{sec:MPS encoding}

Loading classical data into a quantum computer poses a significant challenge in quantum computing~\cite{melnikov2023quantum}. This step is crucial in many quantum algorithms and its implementation for arbitrary states is very complicated and inefficient due to its exponential scaling with the number of qubits~\cite{23/24}.

Nevertheless, preparation of an MPS can be done efficiently, i.e. with linear scaling in the number of qubits~\cite{MPS_preparation}. Various other algorithms are available for MPS preparation, e.g. Refs.~\cite{two_qubits_MPS_encoding, auto_dif_MPS_preparation, gonzalez2023efficient_function_preparation, jumade2023data}, which are quite similar and all work well due to the naturalness of encoding these states into quantum circuits. For a thorough analysis, please see our previous article, Ref.~\cite{melnikov2023quantum}.
In that paper, we also proposed a new vector-encoding scheme based on variational hardware-efficient circuits. 
By simulating these circuits with tensor networks, this scheme enables variational gates to be optimized according to calculations of Riemannian gradients. This provides considerably better accuracy and speed. For example, we numerically simulated how vectors sampled from analytical functions (such as polynomial, trigonometric, and Gaussian distributions) could be efficiently encoded into shallow quantum circuits for up to 100 qubits.

\begin{figure}[h!]
    \centering
    \includegraphics[scale = 0.35]{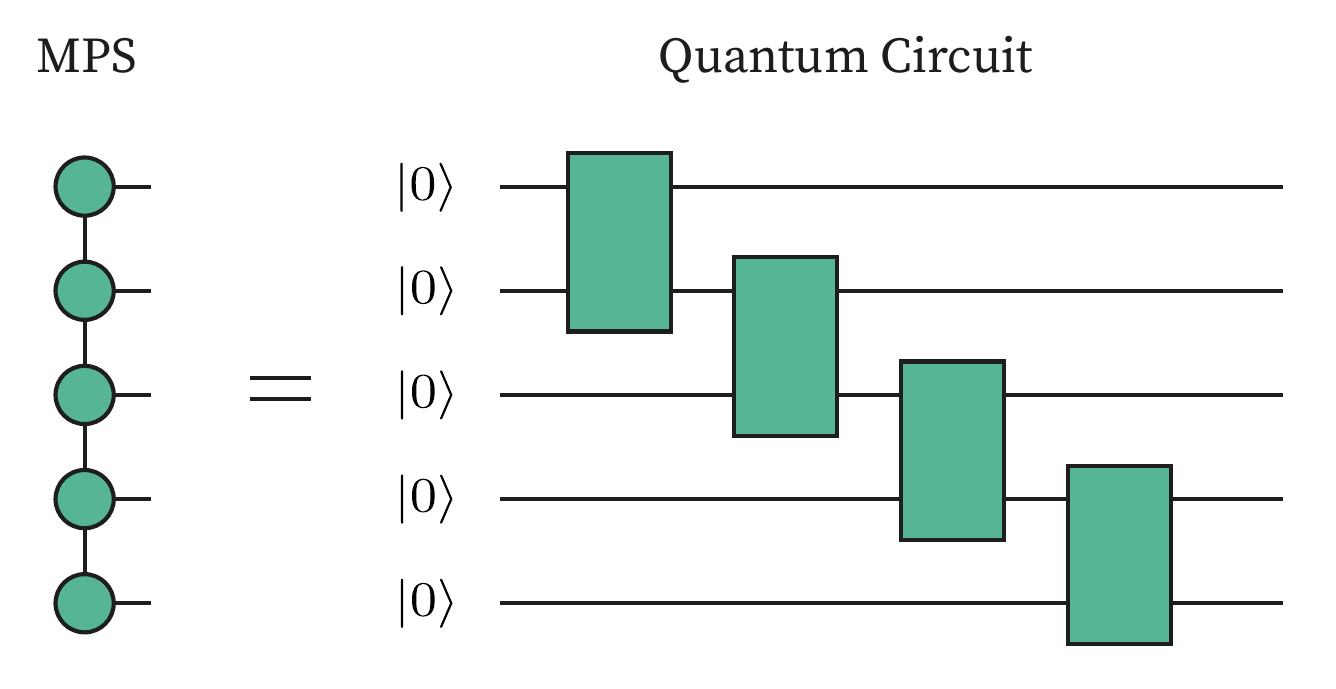}
    \caption{The quantum circuit for encoding an MPS of rank~$2$ involves the sequential application of two-qubit operations. This scheme is extendable to the case of arbitrary rank.}
    \label{fig:mps_enc}
\end{figure}

While much of the research focuses on the approximate preparation of a desired quantum state, a quantum scheme that exactly prepares an arbitrary MPS of rank $r$ is presented in Ref.~\cite{MPS_preparation}. According to this work, any MPS of rank $r$ can be encoded into a step quantum circuit (as shown in Fig.~\ref{fig:mps_enc}), where each gate acts on $[\log(r)] + 1$ qubits, where $[ \cdot ]$ is the ceiling function. To create such a circuit, the MPS cores are first transformed using QR decomposition into isometric cores \cite{malz2024computational}. These isometries are then extended to unitaries projected on zero-state vectors. The resulting network of unitary matrices and zero states form the quantum circuit for preparing the MPS.

\section{Matrix-encoding via Matrix Product Operators}\label{sec: MPO_encoding}

\begin{figure}[ht]
    \centering
    \includegraphics[width = 1\linewidth]{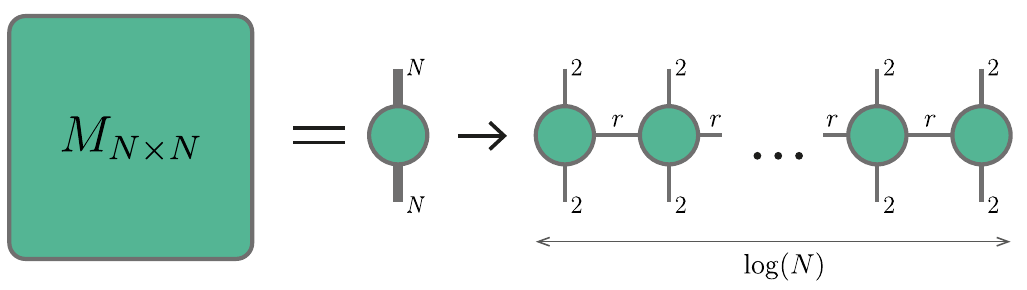}
    \caption{The representation of an $N \times N$ matrix, $M$, as an MPO of rank $r$. This representation occupies $O(\log_2 (N) r^2)$ memory, which is significantly less than the initially required $N^2$ memory in the case of small rank $r$.}
    \label{fig:MPO}
\end{figure}

As depicted in Fig.~\ref{fig:MPO}, an MPO is a representation of a matrix in TT format. Since an MPS can be prepared quite effectively on a quantum computer, the question arises about the same procedure for an MPO. Only one publication has recently emerged dedicated to this particular topic~\cite{nibbi2023block} (see Sec.\ref{sec:conclusion} for a discussion). In this section, we propose our algorithm for effectively encoding MPOs into quantum circuits.

\subsection{Method}\label{sec:Method}

\begin{figure}[ht]
    \centering
    \includegraphics[width = \linewidth]{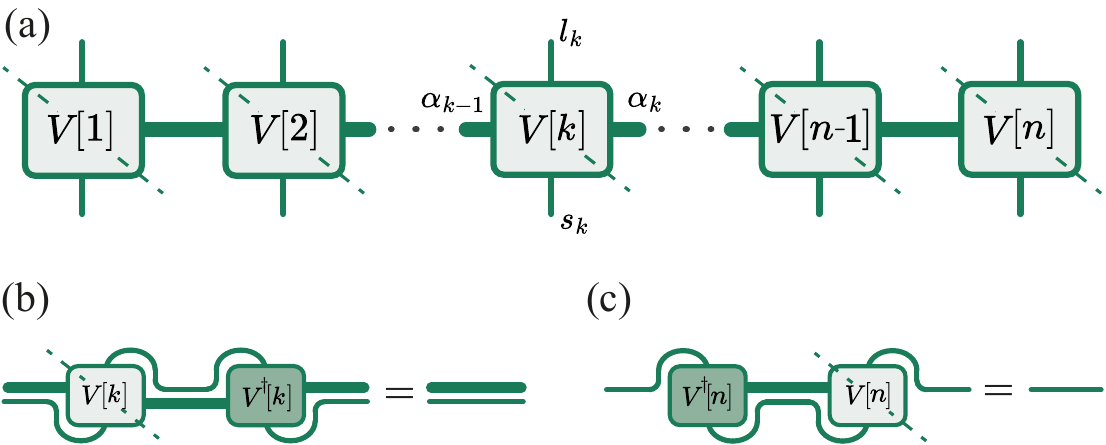}
    \caption{(a) A unitary MPO. 
    The dashed lines indicate the index partition by which the cores should be reshaped to meet the unitarity (isometric) condition.
    After the reshaping, each tensor $V[k]^{ s_k l_k}_{\alpha_{k - 1} \alpha_k}$ is a matrix $V[k]_{\left(l_k \alpha_k \right),( \alpha_{k-1} s_k)}$ with row multi-index $(l_k \alpha_k)$ and column multi-index $(\alpha_{k-1} s_k)$. 
    The unitarity condition for inner cores then reads as $V^*[k]_{( \alpha_{k-1}' s_k'),(l_k \alpha_k)} V[k]_{\left(l_k \alpha_k \right),( \alpha_{k-1} s_k)} = \delta_{\alpha_{k-1}'\alpha_{k-1}} \delta_{s_k' s_k}$, where the asterisk denotes complex conjugation and $\delta_{j j^\prime}$ is the Kronecker delta. 
    For simplicity of notation, we denote this unitarity as condition $V^\dag[k] V[k] = I$, meaning the matrix form of $V[k]$. 
    (b)  The graphical notation of the unitarity condition $V^\dag[k] V[k] = I$ for one of the inner cores. For inner cores, the matrix forms $V[k]$ ($k$ = 2, \dots, n-1) are unitary matrices, whereas $V[1]$ and $V^\dag[n]$ are isometries and satisfy the conditions $V^\dag[1] V[1] = I$ and $V[n] V^\dag[n] = I$, respectively.
    (c)  Graphical notation for the isometric condition $V[n] V^\dag[n] = I$. 
    The right-hand sides of the equalities in (b) and (c) are the identity matrices.}
    \label{fig:unitaryMPO}
\end{figure}

The main step in the exact preparation of an MPS is its orthogonalization. However, for MPOs, there is no such algorithm; it is even not clear whether an arbitrary MPO can be decomposed into an MPO with unitary and isometric cores:
\begin{equation}
    A = A_{s_1 s_2 \dots s_n}^{l_1 l_2 \dots l_n} =
    V\text{\small{[1]}}^{ s_1 l_1}_{\alpha_0 \alpha_1} V\text{\small{[2]}}^{ s_2 l_2}_{\alpha_1 \alpha_2} \dots V\text{\small{[n]}}^{ s_n l_n}_{\alpha_{n - 1} \alpha_n}, 
\end{equation}
where $\alpha_0=\alpha_n=1$. Here we refer to $V[k]$ as a unitary (isometric) core, signifying that it constitutes a unitary (isometric) matrix upon reshaping, aligning with the index division shown in Fig.~\ref{fig:unitaryMPO}.
For inner cores, this unitarity condition reads as
\begin{equation}\label{eq:unitarity_condition}
    V^*[k]_{( \alpha_{k-1}' s_k'),(l_k \alpha_k)} V[k]_{\left(l_k \alpha_k \right),( \alpha_{k-1} s_k)} = \delta_{\alpha_{k-1}'\alpha_{k-1}} \delta_{s_k' s_k},
\end{equation}
where the asterisk denotes complex conjugation and $\delta_{j j^\prime}$ is  the Kronecker delta. To simplify the notation, we denote the unitarity condition Eq.~\eqref{eq:unitarity_condition} as $V^\dag[k] V[k] = I$, meaning the matrix form of $V[k]$. Two boundary cores after reshaping form non-square isometric matrices $V[1]$ and $V^\dag[n]$, which satisfy the isometric conditions $V^\dag [1] V[1] = I$ and $V[n] V^\dag [n] = I$, respectively.
We refer to an MPO with unitary and isometric cores as a unitary MPO.

\begin{figure*}[htb]
        \centering \includegraphics[width=0.9\linewidth]{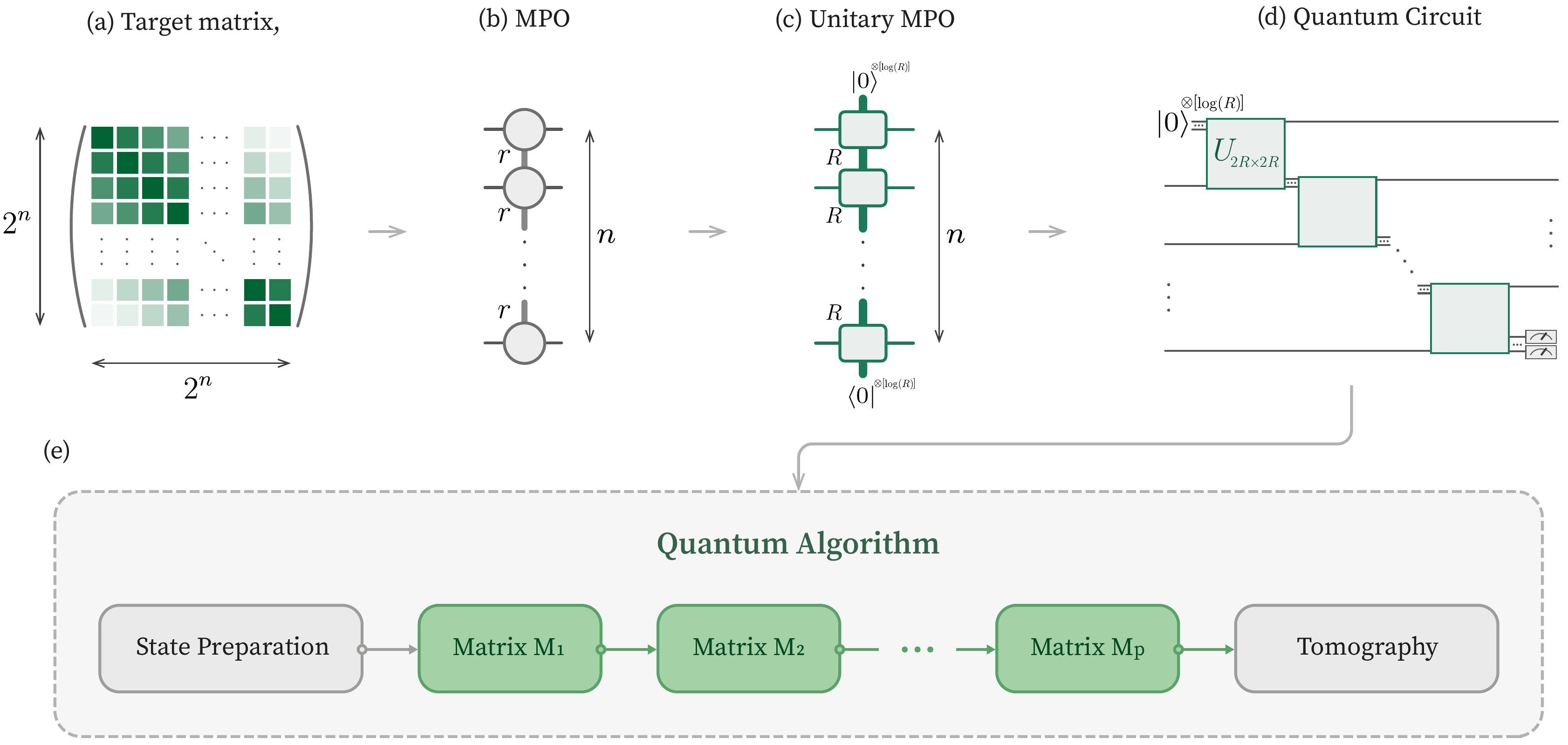}
        \caption{The proposed matrix-encoding algorithm. (a) A target $2^n \times 2^n $ matrix to be encoded. (b) Its  rank-$r$ MPO representation with $n$ cores. (c) The higher rank-$R$ MPO approximation with unitary cores (unitary MPO) achieved via Riemannian optimization. (d) The quantum circuit obtained straightforwardly from the unitary MPO using auxiliary qubits and measurements. (e) Overall, the resulting quantum circuit encodes the target matrix and can be implemented in any quantum algorithm. The matrices that are responsible for operations in the algorithm are denoted  $M_1, M_2,\ldots, M_p$. The simplest example is when the algorithm is a series of matrix-vector multiplications.
        }
        \label{fig_gen_enc_scheme}
\end{figure*}

Our main idea is to combine two key MPS-preparation methods: orthogonalization and a variational approach. This lets us find an MPO with unitary cores that is close to a given MPO such that encoding this unitary MPO into a quantum circuit will be straightforward. All the steps of our algorithm are illustrated in Fig.~\ref{fig_gen_enc_scheme}.

In the first step, we represent the target matrix $M$ in terms of an MPO with rank $r$ before running our encoding algorithm. Next, we approximate the resulting MPO $M_\mathrm{mpo}$ with a unitary MPO of a higher rank, $R$. To achieve this approximation, we employ a manifold gradient descent method, optimizing the error in the Riemannian manifold of unitary cores~\cite{RO_of_isometric_TN, Riemannian_opt}. It's important to note that to obtain unitary cores at the boundaries we artificially utilize ancillas that effectively transform isometric cores into unitary ones, and for such extension, we only need $[\log_2(R)]$ ancillas.

Subsequently, the resulting unitary cores are utilized in a quantum circuit as multi-qubit gates, which are applied sequentially to the system and ancilla qubits. After measuring the ancillas and post-selecting the outcomes, the resulting quantum circuit implements the desired matrix. It is worth emphasizing that our method allows for the implementation of arbitrary, not necessarily unitary, operations, thanks to the post-selection subroutine. Furthermore, the resulting quantum circuit can serve as a building block for any quantum algorithm that requires matrix-vector multiplication by an MPO.

\begin{figure*}[ht]
    \centering
    \includegraphics[width=\linewidth]{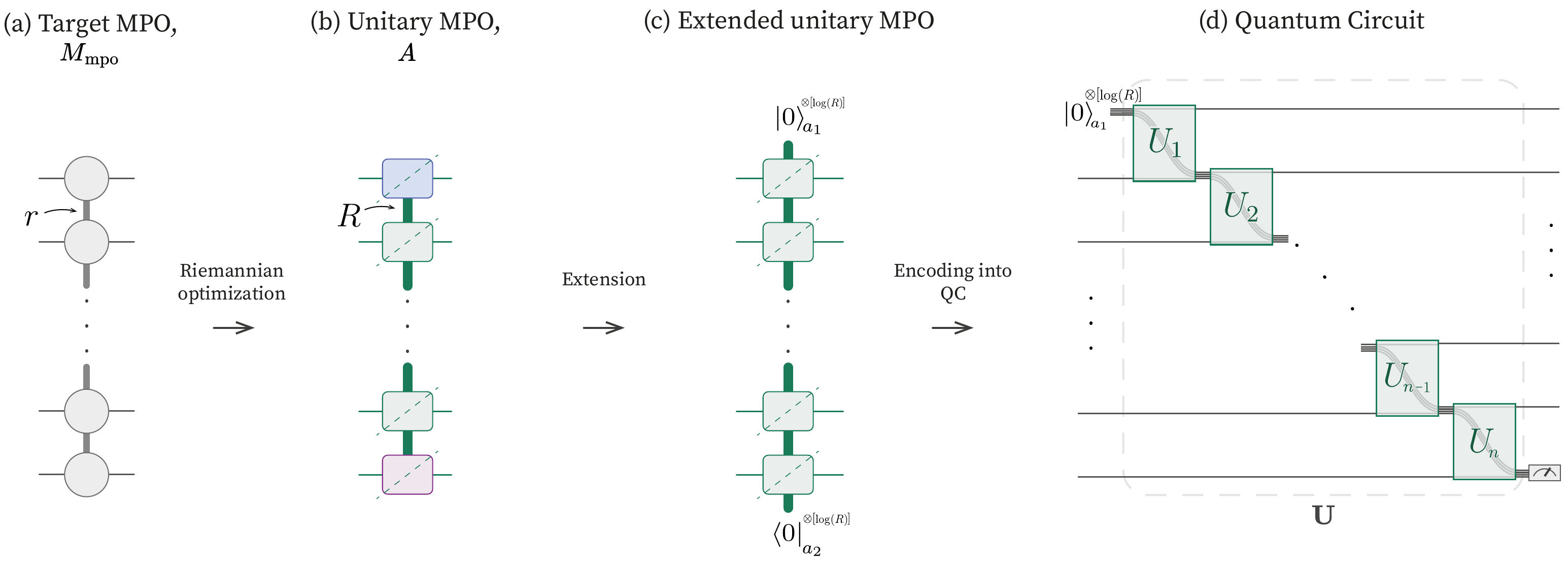}
    \caption{The scheme of encoding an MPO into a quantum circuit. (a) The matrix to be encoded is represented as an MPO of rank $r$. (b) This rank-$r$ MPO is approximated by a higher rank-$R$ unitary MPO using Riemannian optimization. The green dashed lines indicate the index partition by which the cores must be reshaped to form unitary matrices $U_k$ for $k = 2,\dots,n-1$, and isometric matrices for $k = 1,n$ (see Fig.~\ref{fig:unitaryMPO}).
    (c) Two boundary isometric cores are extended to unitary ones, $U_1$ and $U_{n}$. (d) The quantum scheme encoding the target MPO is then produced by sequentially applying unitary gates $U_1,..., U_{n}$ to all system qubits and ancillary qubits in zero state, followed by measurement of the ancillas and post-selection.}
    \label{fig:method}
\end{figure*}

Since any matrix can be approximated by an MPO with some error, our algorithm can approximately encode any matrix into a quantum computer. In addition, in many applications, such as partial differential equations, quantum physics, or quantum chemistry, an analytical expression of a matrix in MPO format is readily available~\cite{khor-low-rank-kron-P2-2006,grasedyck-kron-2004,sdwk-nmr-2014}.

\subsubsection{Preprocessing of matrices using the unitary MPO}

To begin with, it is crucial to address a significant aspect of the normalization of the matrix we want to encode. When implementing the action of a matrix, $M$, on a state represented by a density matrix, $\rho$, on a quantum computer, the resulting transformed state becomes $M \rho M^\dag$. However, it is essential to consider the constraint $\mathrm{Tr} [M \rho M^\dag ] \leq 1$, as quantum operations do not increase the trace~\cite{nielsen2002quantum_computing}. This constraint implies that the absolute values of the elements of matrix $M$ must not exceed a certain upper bound.

To overcome this limitation, we introduce a real normalization constant, $c$, and our objective shifts towards realizing the $c^{-1} M$ matrix on the quantum computer instead of just $M$. From a practical standpoint, this implies the necessity of remembering this normalization factor and incorporating it into the final calculations. By incorporating the normalization constant, $c$, we ensure that the quantum computer can efficiently handle the transformation $c^{-2} M \rho M^\dag$, satisfying the trace constraint while preserving the desired operation. It's important to emphasize that considering this normalization factor is merely a technical requirement, which can be straightforwardly managed in the computations.

As mentioned above, our first step is to find a unitary MPO, $A$, and a normalization constant, $c$, such that $c A$ approximates $M_\mathrm{mpo}$ as closely as possible.
The TT-rank, $R$, of $A$ should be taken to be larger than the TT-rank, $r$, of $M_\mathrm{mpo}$ since unitarity takes away some of the degrees of freedom. We formulate this approximation task in terms of finding the minimum of the cost function:
\begin{align*}
    \min_{c, A} \hspace{10pt} & \hbox
    {C} = ||c A - M_\mathrm{mpo}||^2  \\
        \text{subject to } \hspace{6pt} &V[k] : V^\dag[k] V[k] = I \mbox{ } \forall k = 1,\dots,n-1, \\
        &  \hspace{28pt} V[n]V^\dag[n] = I,
\end{align*}
where $||\cdot||$ denotes the Frobenius norm. Here, the optimum is taken over both $c$ and $A$. We solve this optimization problem by alternating descent, and the optimization algorithm (steps (a)-(b) in Fig.~\ref{fig:method}) reads as follows:
\begin{enumerate}
    \item [Step 0.] Fix an initial approximation $A$.
    \item [Step 1.] Find $c$ that minimizes the cost function $||c A - M_\mathrm{mpo}||^2$ with the fixed $A$ via formula 
    \begin{equation} \label{step2}
        c = \Re\frac{\mathrm{Tr}[ A^\dag M_\mathrm{mpo} ] }{||A||^2}.
    \end{equation}
    \item [Step 2.] Find a new approximation $A$ via one step of the Riemannian gradient descent for the problem
    \begin{subequations}
    \begin{align} 
    \min\limits_{A} \hspace{10pt} &C = ||c A - M_\mathrm{mpo}||^2 \label{cost} \\
        \text{subject to } \hspace{2pt} &V[k] : V^\dag[k] V[k] = I \mbox{ } \forall k = 1,\dots,n-1, \label{condition}\\
        &  \hspace{28pt} V[n]V^\dag[n] = I , \label{condition_}
            \end{align}
    \end{subequations}
    and go to Step 1.        
\end{enumerate}

Step 2 is the most challenging because it involves optimization with constraints (Eq.~\eqref{condition},~\eqref{condition_}) that ensure the isometry of the MPO cores. Any $m \times p$  $(m \geq p)$ isometric matrix belongs to the Stiefel manifold $\text{St}(m,p,\mathbb{C}) := \{ V \in \mathbb{C}^{m \times p} | V^\dag V = I \}$, which is a Riemannian manifold. The cost function Eq.~\eqref{cost} is thus defined on the Cartesian product of $n$ Stiefel manifolds. To find a set of isometric matrices $V[k]$ minimizing this cost function, one can use gradient-based optimization methods, such as Riemannian gradient descent which works for optimization problems with constraints~\cite{QGOpt}.

In regular gradient-based methods, such as gradient descent, the variables being optimized are updated by moving in the opposite direction to the cost function gradient: $ V[k] \rightarrow V[k] - \alpha \partial_k C $, where $\alpha$ is a step size and $\partial_k C = \partial C / \partial V[k]$. In the Riemannian generalization of these methods, to determine the direction of this movement, instead of $\partial_k C $, its projection $\partial_k^R C \in \mathcal{T}_{V[k]}$ onto the space tangent to the Stiefel manifold at point $V[k]$ is used. $\partial_k^R$ is called the Riemannian gradient, and to update the values of matrices $V[k]$, a retraction onto the manifold is used: $V[k] \rightarrow \mathcal{R}_{V[k]}(- \alpha \partial_k^R C)$~\cite{RO_of_isometric_TN, Riemannian_opt}. Besides conventional gradient descent, there are Riemannian generalizations of other, more advanced optimization methods, such as the adaptive moment estimation algorithm (ADAM)~\cite{adam, Bcigneul2018RiemannianAO} that we use in this work.

To sum up, Riemannian gradient descent is an iterative process where, at each step, the algorithm calculates the Riemannian gradient of the current point's cost function. Subsequently, it employs the retraction in the opposite direction to determine the next point in the optimization procedure. Since retraction along the manifold is employed, the updated optimization point inherently belongs to the manifold. In our case, this means that conditions Eq.~\eqref{condition} and ~\eqref{condition_} are automatically satisfied.

Once we get a unitary MPO, $A$, that is the approximation of $c^{-1} M_\mathrm{mpo}$, all the inner cores of $A$ are already unitary matrices (after the corresponding reshape). However, the tensors at the boundaries ($V[1]$ and $V[n]$) are only isometries and they need to be expanded so that they become unitary. Therefore, we extend them to unitary matrices and project them onto $\ket{0}$ states to maintain the correspondence with the matrix, see Fig.~\ref{fig:method}(c). Now all the cores are unitary and the transition to the quantum circuit is straightforward.

\subsubsection{Encoding using quantum circuit}

In the context of quantum circuits, the $\ket{0}_{a_1}$ state (see Fig.~\ref{fig:method} (c)) corresponds to preparing additional ancilla qubits in the zero state. At the same time, the $\ket{0}_{a_2}$ state corresponds to the projection onto the zero states, i.e. measurements with post-selection. Thus, we obtain the following steps for the quantum part:

\begin{enumerate}
    \item Prepare ancillary qubits in the zero state $\ket{0}^{\otimes [\log_2(R)]}$ and leave the other qubits alone since our goal is only to encode the matrix. We will receive an initial state as input to which we will then apply this matrix.
    \item Sequentially apply gates $U_1,... , U_n$, which should be decomposed into one-qubit and two-qubit operations (see below).
    \item Measure the ancillary qubits in the computational basis. Note that one should not measure the qubits that were originally introduced as ancillary, but the qubits from the opposite end; see Fig.~\ref{fig:method} (d).

    If all measurement outcomes are equal to $0$, stop the algorithm. Else go back to step 1.
\end{enumerate}

As stated, the multi-qubit gates $U_1,..., U_n$ must be decomposed into one-qubit and two-qubit operations. Each $U_k$ acts on $([\log_2(R)] +1)$ qubits, where $R$ is the TT-rank of the unitary MPO, $A$. Ref.~\cite{Rakyta2022approaching} proposes a numerical strategy to approximately decompose a general multi-qubit unitary with a CNOT gate count very close to the theoretical lower limit presented in Ref.~\cite{TheorLimQGdec}. According to Ref.~\cite{TheorLimQGdec}, decomposing an arbitrary $m$-qubit gate requires $\simeq 4^{m - 1}$ CNOT gates. In our case, we have to use $\simeq n \cdot 4^{\log_2 R} = n R^2$ CNOT gates in total. Therefore, the lower the rank we take for the approximation, the simpler the quantum circuit turns out, but the worse the accuracy of the approximation.

\subsubsection{Post-selection success probability analysis}\label{sec:post-selection}

The introduction of ancillary qubits makes it possible to implement non-unitary matrices on a quantum computer. 
The post-selection of ancilla measurements makes the encoding algorithm probabilistic. 
In this section, we examine the probability of successful encoding.

The quantum circuit $\mathbf{U}$ (see Fig.~\ref{fig:method}) implements the action of the unitary MPO $A$ only if the ancillaries are all initialized and measured in the state $\ket{0}$. 
We refer to the probability of achieving this successful measurement as the success probability.

The success probability depends on the initial state $\ket{\psi_\mathrm{in}}$ to which the quantum circuit is applied and can be expressed as follows:
\begin{equation}
{\Pr}_\text{suc} =  \mathrm{Tr} \big [ \bra{0}_{a_2}\mathbf{U} \ketbra{\psi_\mathrm{in}} \otimes \ketbra{0}_{a_1} \mathbf{U}^\dag \ket{0}_{a_2} \big ],
\end{equation}
Using $\bra{0}_{a_2} \mathbf{U} \ket{0}_{a_1} = A$, we can express the probability as:
\begin{equation}\label{eq:succes_probability_A}
    {\Pr}_\text{suc} 
 = \mathrm{Tr} \big [A^\dag  A \ketbra{\psi_\mathrm{in}} \big ] = ||A \ket{\psi_\mathrm{in}} ||^2.
\end{equation}

We can represent the matrix $A^\dag A$ in terms of its eigenvalues and eigenstates: $A^\dag A = \sum_k \lambda_k^2(A) \ket{\lambda_k(A)}\bra{\lambda_k(A)}$. 
In this case, the eigenvalues of $A^\dag A$ are the squared singular values $\lambda_k(A)$ of the matrix $A$.
Thus the success probability reads
\begin{equation}\label{eq:succes probability spectrum}
    {\Pr}_\text{suc} 
 = \sum\limits_k \lambda_k^2(A) \left|\braket{\lambda_k(A)}{\psi_\mathrm{in}}\right|^2. 
\end{equation}
Therefore, the success probability depends on the singular value spectrum of $A$, as well as the initial state $\ket{\psi_\mathrm{in}}$ and its overlap with the eigenvectors of $A^\dag A$. 

In certain scenarios, however, the initial state does not affect the success probability. 
This happens when all the singular values are identical. 
One such example is solving partial differential equations (PDEs) in closed quantum systems, where encoded matrices are unitary with singular values all equal to one \cite{rivas2012open}. 

According to Eq.~\eqref{eq:succes probability spectrum}, the maximum success probability for a fixed $A$ is achieved when $\ket{\psi_\mathrm{in}} = \ket{\lambda_{\max}(A)}$:
\begin{equation}\label{eq:max_probability}
    \max\limits_{\psi_\mathrm{in}} {\Pr}_\text{suc} = \lambda_{\max}^2(A) \leq 1,
\end{equation}
which also implies a limitation on the $A$'s singular values: $ \lambda_k(A) \leq 1$ for all $k$.

The above reasoning is general and also applies to other matrix encoding methods that use auxiliary qubits. 
In our algorithm, $A$ is a unitary MPO approximation of the normalized original matrix $M$. 
In the case of an exact approximation, $A = c^{-1}M$ and $\Pr_\mathrm{suc} = c^{-2} \sum_k \lambda_k^2(M) \left|\braket{\lambda_k(M)}{\psi_\mathrm{in}}\right|^2$. 
Then, taking into account Eq.~\eqref{eq:max_probability}, the probability of success for each fixed $\ket{\psi_\mathrm{in}}$ reaches its maximum if $c = \lambda_{\max}(M)$. 

Thus, a good strategy to increase the success probability is to normalize the original matrix $M$ so that its maximum singular value is one, and then approximate it with a unitary MPO for subsequent encoding into a quantum circuit. 
However, finding the maximum singular value in the general case is a difficult task for high-dimensional matrices. 
Therefore, we propose to optimize the normalization constant when finding an approximation of the target matrix by a unitary MPO.

\subsubsection{The average success probability}

As stated in the previous section, the initial state $\ket{\psi_\mathrm{in}}$ to which the quantum circuit is applied can affect the success probability. 
The initial state depends on the particular task. To study the success probability without considering specific initial states, we can uniformly average over all possible initial states of the system. 

The average of uniformly distributed pure $n$-qubit quantum states with respect to the Haar measure \cite{mele2024introduction} equals $\frac{1}{2^n} I$, where $I$ is the identity matrix. Thus, according to the Eq.~\eqref{eq:succes_probability_A} the average success probability reads
\begin{equation}\label{eq:success_prob}
\langle {\Pr}_\text{suc} \rangle = \frac{||A||^2}{2^n},
\end{equation}
where $||\cdot||$ denotes the Frobenius norm. 
This equation shows that the Frobenius norm of the unitary MPO, which depends on its singular value spectrum, determines the average success probability. At first glance, it might seem that $\langle {\Pr}_\text{suc} \rangle$ drops exponentially with the number of qubits. 
However, this is not the general case, since $||A||^2$ can grow exponentially with the number of qubits. 
For instance, if $A$ is a unitary matrix, then $||A||^2 = 2^n$ and the average success probability is $1$. 
In Sec.~\ref{sec:results}, we numerically simulate several commonly encountered matrices and find that there is no exponential decrease in the average success probability.
 
\begin{figure*}[ht]
    \centering
    \includegraphics[width=1.0\linewidth]{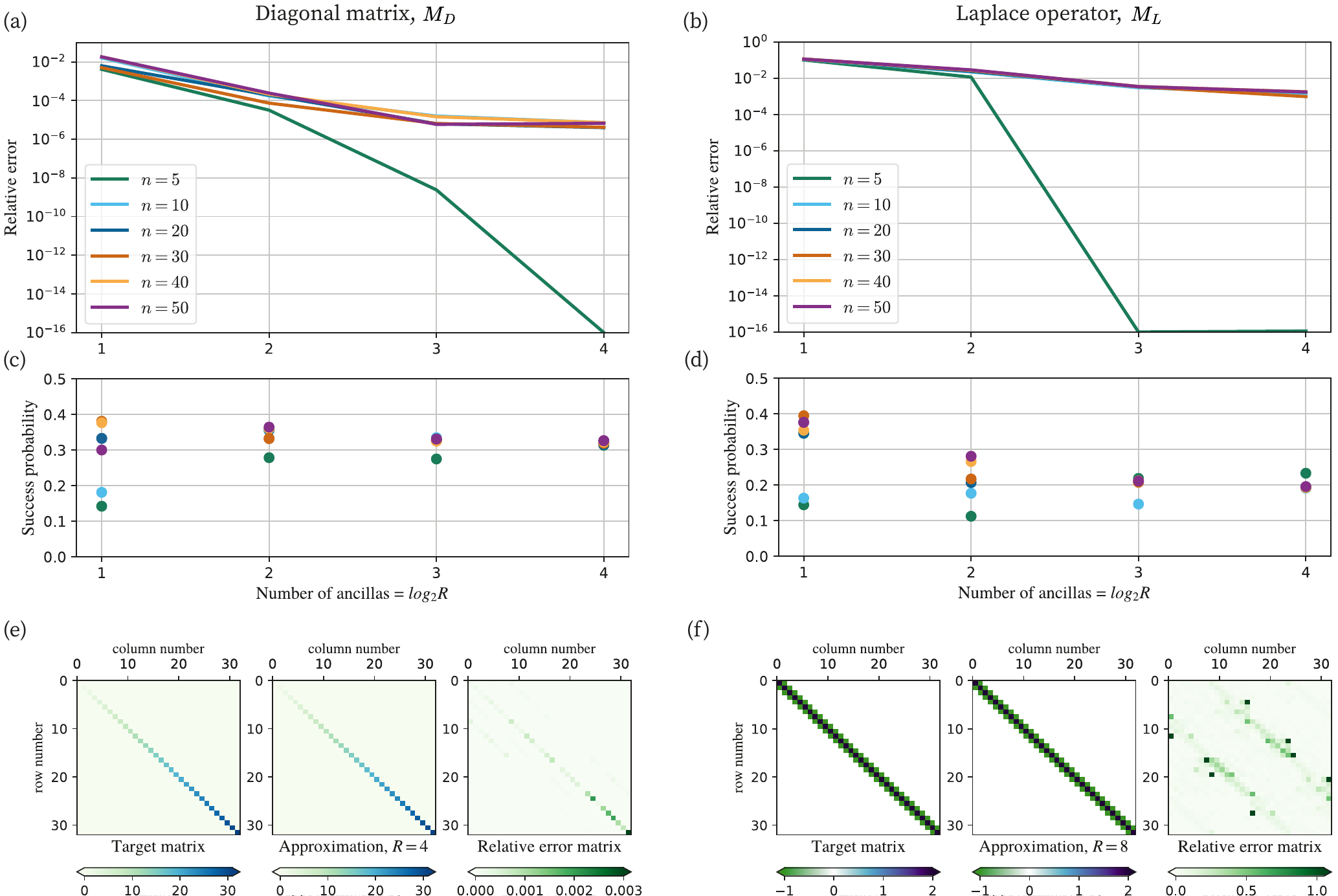}
    \caption{The minimum relative error, $\varepsilon$, achieved in the approximation of (a) the diagonal and (b) the Laplace matrices by unitary MPOs depending on its rank, $R$, for six different matrix sizes $2^n \times 2^n$, where $n$ is the number of qubits. The corresponding average success probabilities for (c) the diagonal and (d) the Laplace matrices. 
    For convenience, we denote the abscissa axis as the number of used ancillas equal to $\log_2 R$. Examples of the matrix approximation for five qubits (i.e., the matrix sizes are $2^5 \times 2^5$) with errors $3 \cdot 10^{-5}$ for (e) the diagonal matrix and $2 \cdot 10^{-15}$ for (f) the Laplace matrix. The relative error matrix is calculated using the formula $||M||^{-1}|cA - M|$.}
    \label{fig:errs_diag_lap}
\end{figure*}

\subsection{Results}\label{sec:results}

\subsubsection{Diagonal and Laplace matrices}

To demonstrate the capabilities of our method, we consider matrices that are frequently used in computational mathematics and encode them into multi-qubit quantum circuits. In this work, we consider a matrix, $M_D$, with a discretized linear function on the diagonal and the Laplace operator, $M_L$, corresponding to the second derivative matrix after finite difference discretization~\cite{fin_dif}, both of which are not unitary:
\begin{equation}
    M_D = 
    \begin{pmatrix}
    0 \\
    & 1 \\
    & & \ddots \\
    & & & 2^n -1 
    \end{pmatrix} ; \hspace{5pt}
    M_L = 
    \begin{pmatrix}
    2 & -1 \\
    -1 & 2 & \ddots \\
    & \ddots & \ddots & -1 \\
    & & - 1& 2 
    \end{pmatrix} .
\end{equation}

The Laplace operator, $M_L$, is necessary for the solution of PDEs~\cite{Poisson_variational, fast_poisson_circuit} while the diagonal matrix, $M_D$, as we describe in the Applications Sec.~\ref{sec:Applications}, can be used in both optimization~\cite{grasedyck2019finding_extreme_diagonal} and PDEs. For any $n > 1$, the MPO forms of the $M_D$ and $M_L$ matrices have TT-ranks $r$ equal to 2 and 3, respectively. To investigate the performance of our method, we conducted the following experiment for both matrices: fixing the number of qubits, $n$, and the rank of the unitary MPO, $R$, we estimate the maximum achievable approximation accuracy, $\varepsilon = ||cA - M||^2 / ||M||^2$, and corresponding average success probability defined in Eq~\eqref{eq:success_prob}.

The results of this experiment are presented in Fig.~\ref{fig:errs_diag_lap} -- the minimum relative error achieved during the approximation of these matrices by the unitary MPOs is shown in the top plots and the corresponding success probability is shown in the middle plots. For both matrices, our method can achieve reasonable accuracy (i.e., errors of about $0.01\%$ and $0.5\%$ for $M_D$ and $M_L$, respectively) and a sufficient success probability (i.e., more than $10-20\%$ up to 50 qubits).

Importantly, the error and the success probability almost do not depend on the number of qubits. In the bottom part of Fig.~\ref{fig:errs_diag_lap}, we show examples of the resulting approximations and their comparison with exact matrices.

\subsubsection{Approximation of multi-controlled Toffoli gate}

Our scheme can be also applied to decompose multi-controlled quantum gates, which are widely used in quantum computing. As an example, in this work we consider the decomposition of multi-controlled Toffoli gate (MCT), a crucial matrix in quantum computing that finds application in various quantum algorithms, including Grover's algorithm~\cite{grover1996fast}, Shor's factorization algorithm~\cite{shor1999factorization}, and solving a linear system of equations~\cite{HHL, childs2017quantum_linear}. Therefore, the efficient decomposition of this gate is one of the key tasks of quantum computing~\cite{shende2008_cnot_toffoli, saha2020_toffoli_for_grover, rqc_toffoli}. 

The MCT gate for $n$ qubits can be represented in matrix form as follows:
\begin{equation}\label{MCT_matrix}
    M_\text{MCT} = 
    \begin{pmatrix}
    1 \\
    &  \ddots \\
    &  & 1 \\
    &  & &  0 & 1 \\
    &  & &  1 & 0 \\
    \end{pmatrix}.
\end{equation}

\begin{figure}[ht]
    \centering
    \includegraphics[width=1.0\linewidth]{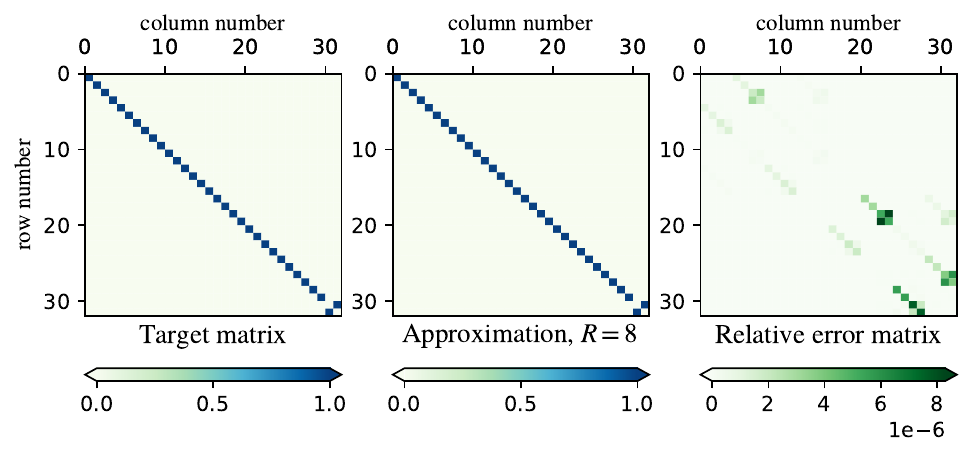}
    \caption{Example of the multi-controlled Toffoli gate approximation by a unitary MPO for $5$ qubits with an error $\varepsilon = 6 \cdot 10^{-10}$. The relative error matrix is calculated using the formula $||M||^{-1}|cA - M|$, where $M$ is a MCT gate matrix \eqref{MCT_matrix}.}
    \label{fig:MCT}
\end{figure}

\noindent The important feature of this matrix is that its TT-rank in the MPO representation equals $2$, which allows us to successfully employ our encoding method. It is possible to approximate the MCT gate with high accuracy by a unitary MPO with sufficient rank $R$, achieving an error $\varepsilon$ less than $10^{-9}$. An illustrative example of such an approximation for a $5$-qubit system is presented in Fig.~\ref{fig:MCT}.

\begin{figure}[ht]
    \centering
    \includegraphics[width=1.0\linewidth]{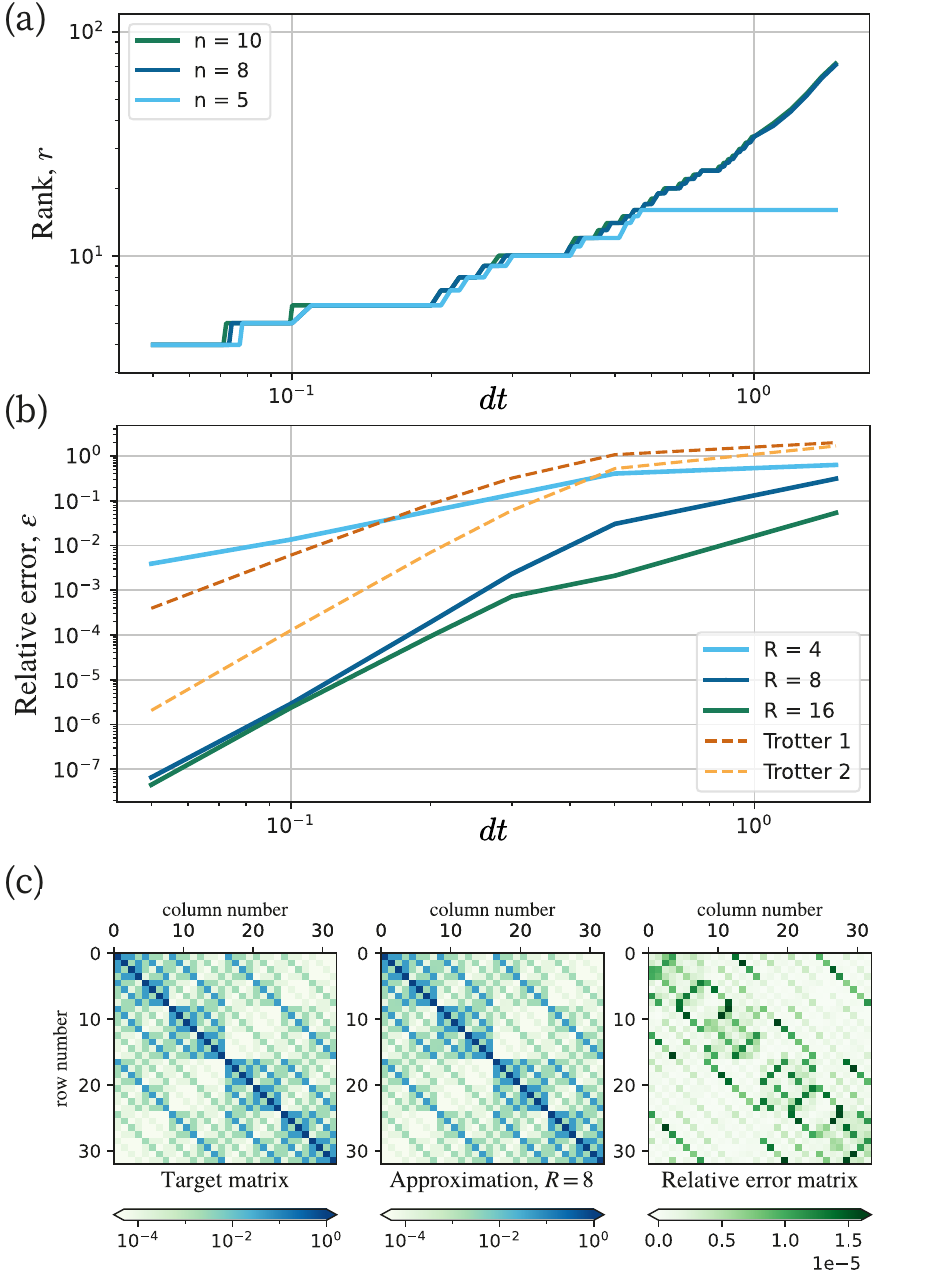}
    \caption{(a) The dependence of the MPO representation rank $r$ of the matrix $e^{-i H dt}$ with respect to $dt$. At small $dt$, the rank $r$ weakly depends on the number of qubits $n$. However, as $dt$ increases, the rank $r$ plateaus and reaches a maximum value that is solely based on the number of qubits and equals $4^{[\frac{n}{2}]}$. The rounding tolerance for the MPO approximation is $10^{-6}$. (b) The relationship between $dt$ and the minimum achieved error $\varepsilon$ in the approximation of $e^{-i H dt}$ by a unitary MPO of rank $R$ for 3 different values of $R$ (solid) and by Trotter product formulae of 1st and 2nd order (dashed). The number of qubits equals $8$ in all cases. (c) An example of the evolution operator approximation for $5$ qubits and $t = 0.05$ with an error $2 \cdot 10^{-8}$. The relative error matrix is calculated using the formula $||M||^{-1}|cA - M|$.}
    \label{fig:hamiltonian}
\end{figure}

\subsubsection{Hamiltonian simulation}

One of the primary areas of use for quantum computation is the simulation of Hamiltonian dynamics. The challenge of simulating the behavior of quantum systems was the original inspiration for quantum computers, and it still serves as one of their major potential applications. Hamiltonian simulation acts as the fundamental approach for simulating quantum systems and is also applicable to quantum algorithms and quantum chemistry.

To demonstrate how well our algorithm can perform in approximating an evolution operator, we pick an open quantum Ising chain with both transverse and longitudinal fields for the Hamiltonian:
\begin{equation}
    H = J \sum\limits_{k = 1}^{n - 1} Z_k Z_{k + 1} + g \sum\limits_{k = 1}^{n} X_k + h \sum\limits_{k = 1}^{n} Z_k .
\end{equation}
Here, $X$ and $Z$ are Pauli matrices and a non-integrable point is chosen for the Hamiltonian parameters:  $J = 2$, $g = 1$, and $h = 1$. 

Our task is to encode the evolution operator $e^{-i H dt}$ into a quantum circuit with a small value of $dt$, for the evolution operator to have small ranks; see Fig.~\ref{fig:hamiltonian}(a). Then, by repeating this circuit several times, we can perform the evolution at any desired time $T$. In Fig.~\ref{fig:hamiltonian}(b), we present the results of the comparison in terms of the accuracy of our algorithm with the Trotter decomposition that was used in a recent quantum utility work~\cite{kim2023evidence_IBM_advantage}. It can be seen that our algorithm demonstrates better accuracy when choosing a sufficient rank $R$ at reasonable $dt$ values.

\section{Readout via MPS tomography}\label{sec:TTDE}

The last important step of any quantum algorithm is a readout of the output state. In general, to implement efficient tomography of $n$-qubit quantum states, it is necessary to perform an exponential (in $n$) number of measurements, and an exponential amount of classical memory and computing power is required. To overcome this, various approaches have been introduced: classical shadow~\cite{huang2020predicting}, neural network methods~\cite{Carrasquilla2019}, and TNs~\cite{cramer2010_tomograthy, eisert2013entanglement, pastaq}.
Here, we propose to use TN-based tomography~\cite{cramer2010_tomograthy} for several reasons:
\begin{itemize}
    \item It provides a more comprehensive representation of the vector in the MPS format, extending beyond the mere capability to measure observables~\cite{huang2020predicting}.
    \item Having the output in the TN format serves as a valuable input for subsequent stages of a hybrid algorithm. Our scheme, for instance, involves the progression from MPS to quantum computing and back to MPS.
    \item TNs offer a more transparent and explainable approach compared to alternatives such as neural networks~\cite{khrulkov2017expressive, tt-cross}
\end{itemize}

\subsection{Method}

In our experiments, we employed MPS tomography as outlined in Ref.~\cite{pastaq}. The algorithm can be summarized as follows: 
\begin{itemize}
    \item Firstly, choose the initial MPS approximation with fixed rank $r$ for the output state. Note that with this method, complex-valued vectors can also be measured. 
    \item Secondly, measure the output state in different randomly chosen bases (for example, XXYZ...ZXY), where the outcomes, $x_i$, in bases, $b$, can be denoted as $x_i^{(b)}$.
    \item Next, introduce a loss function to quantify the relative entropy between the predictions from the MPS approximation and the actual measured probabilities.

    \begin{equation}
    \text{Loss} = - \frac{1}{N} \sum_{i=1}^{N} \log P(x_i^{(b)}).
    \end{equation}

    Here, $P(x_i^{(b)})$ represents the probability of obtaining the measured outcome $x_i$ in basis $b$, with $N$ denoting the number of measurements. We obtain this probability by projecting the MPS at the current iteration onto the corresponding basis vector, which is done effectively in the format of tensor networks.
    \item Subsequently, employ stochastic gradient descent on the parameters of the MPS cores to minimize the loss as much as possible~\cite{SGD}.

\end{itemize}

\subsection{Results}
The key aspect of the above method lies in its ability to maintain a linear scaling of the number of measurements (or samples) with the number of qubits for tomographing a state while ensuring a specified level of accuracy. 
To explore this, we examine the MPS tomography of a variational quantum circuit (VQC) output state, typically well-described in the form of MPS when the number of layers is limited~\cite{perelshtein2023nisq}.

\begin{figure}[ht]
\centering
\includegraphics[width=0.95\linewidth]{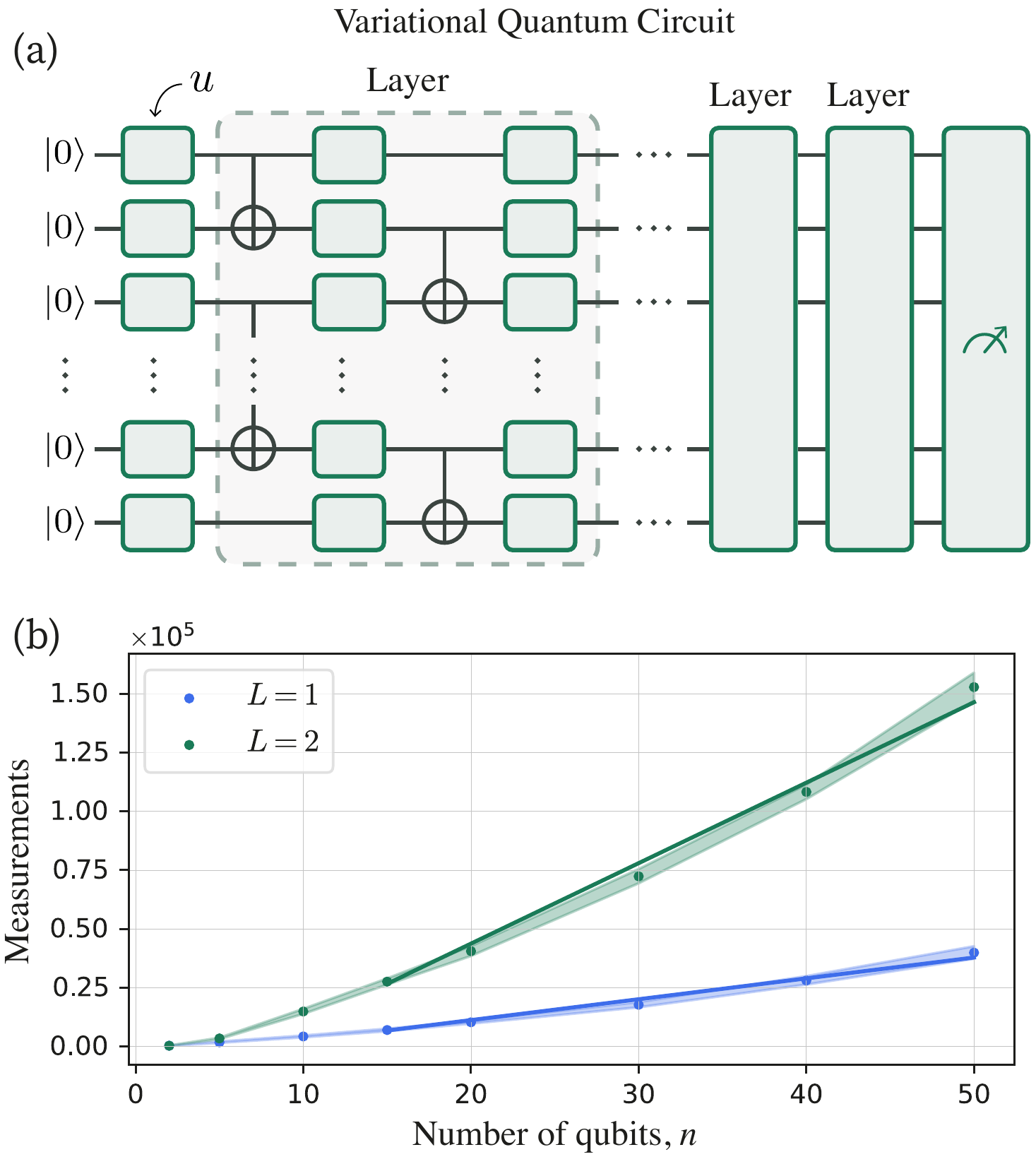} 
    \caption{(a) A VQC example. The circuit is composed of multiple layers, while each layer is a sequence of non-parametric CNOT gates and single-qubit variational gates $u$. The parameters of the variational gates $u$ are determined at random for our experiment. (b) MPS tomography results for the output states of random VQCs. The plot shows the number of measurements required to obtain 99\% quantum fidelity, depending on the number of qubits $n$ in the circuit with a fixed numbers of layers, $L$.}
    \label{fig:varqc_plot}
\end{figure}

Fig.~\ref{fig:varqc_plot} depicts an example of a variational quantum circuit and the results of the MPS tomography for such a circuit. 
Notably, we observe a linear correlation between the number of measurements and the number of qubits when the number of layers is fixed.

\section{Applications}\label{sec:Applications}

At first glance, it may seem that this framework does not find a lot of applications. However, in this section, we consider several examples. In essence, what a quantum computer can do is sequentially perform {\tt Matvec} operations:
\[X = U_n U_{n - 1}  \dots U_{2} U_1 X_0    ,\]
where $X_0$ is the initial state-vector and $U_i$ are  matrices (not necessarily unitary). Therefore, regarding quantum algorithms, encoding matrices into quantum circuits is one of the most important tasks. This is precisely the task that we provide a solution to in this article - the main feature is that the number of gates in the proposed circuit has a linear scaling with the number of qubits and it is possible to control the size of the circuit, varying the accuracy of the approximation. In contrast, the encoding of an arbitrary $n$-qubit matrix scales exponentially~\cite{TheorLimQGdec}.

To see how a quantum implementation can possess advantages over a classical one, consider the following. If the task is to implement a matrix-vector multiplication, $Ax$, then the classical realization via tensor networks has complexity $O(n d r^{3}(A) r^{3}(x))$, where $n^d$ is the vector size, and $r(A)$ and $r(x)$ are TT-ranks of $A$ and $x$, respectively. The complexity of the quantum implementation does not depend on the rank $r(x)$, only on the rank $r(A)$. Therefore, when there are large solution ranks and small operation ranks, the quantum computer provides advantages. As will be shown below, this approach has numerous applications, including the solution of PDEs, optimization, machine learning, and quantum chemistry.

\subsection{Simulation}

When almost any differential equation is discretized in time and space, it is reduced to an iterative scheme, which in its simplest form is written as follows:
\[x_{i + 1} = M x_{i} ,\]
where $M$ represents the derivative operator. This matrix $M$ is usually diagonal or contains first and second derivatives, and is therefore well-represented by an MPO with small TT-ranks~\cite{oseledets2011tensor, kazeev2012_toeplitz, kornev2023_NS}. Often matrix $M$ does not depend on the iteration number, $i$, and is fixed. In this case, the implementation of our numerical scheme on a quantum computer is advantageous because it can be realized in a way that is independent of the ranks of the solution, $x_i$, during the iterations. Consequently, increasing the ranks during the iterations, which is often the case in such problems~\cite{kornev2023_NS}, does not affect the computational complexity on a quantum computer - it will only depend on the efficiency of encoding matrix $M$ into a quantum circuit.

For example, if we want to solve the heat-diffusion equation:
\[
\begin{aligned}
\frac{\partial u}{\partial t} & = k^2 \Delta u , \quad 
x \in [0, 1], t \geq 0; \\
u(x, 0) & = u_0(x), \quad x \in [0, 1]; \\
u(0, t) & = u(1, t) = 0, \quad t \geq 0;
\end{aligned}
\]
after discretization (explicit in time), it is reduced to 
\[
\begin{aligned}
u_{i + 1} & = (I + \frac{\Delta t}{\Delta x^2} k^2 M_L) u_i,  \\
u_0 & = \hat{u}_0,
\end{aligned}
\]
where $M_L$ is the Laplace operator from Section~\ref{sec:results}.  
Here, $\Delta t$ and $\Delta x$ are discretization steps in time and space, respectively, $u_i$ is the discretized solution at time $i \Delta t$. If the initial ranks of $u_0$ are large then this scheme is better suited to a quantum implementation than a classical one.

\subsection{Optimization}

Tensor trains are used to tackle black-box optimization problems~\cite{sagingalieva2023hybrid, belokonev2023optimization, morozov2023protein}.
Remarkably, it is often possible to reduce an optimization problem to the task of finding the maximum element in a tensor $y$, regardless of whether it is a discrete, continuous, or black box optimization (including minimization) task~\cite{batsheva2023protes}. 
In this formulation, one can always apply the so-called Power method~\cite{chertkov2022optimization_power}. 
That is, finding the maximal element in the normalized tensor $y^n$ instead of $y$, which is much easier since the largest element in $y^n$ is much larger, in relative terms, compared to the largest element of $y$. Ideally, with a sufficiently large degree, $n$, the normalized $y^n$ turns into a delta function in a peak with a maximum of $y$. 

We can approximate $y$ as a TT, $y_{TT}$. The overall accuracy of this approximation can be poor as long as the maximal element of $y_{TT}$ is close to the maximal element of $y$. We can now create a quantum circuit that realizes a diagonal MPO matrix, $M_y$, that corresponds to $y_{TT}$ using our algorithm. Thus, we can implement the Power method on a quantum computer, preparing the initial state as a vector of units:

\[y^n \simeq   M_y M_y  ... M_y  M_y \frac{1}{\sqrt{N}}\begin{pmatrix}
    1\\
    1\\
    \vdots \\
    1
\end{pmatrix}. \]

It is important to note that when implemented on a quantum computer, the normalization will be automatically observed, due to post-selection (see Sec.~\ref{sec:post-selection}). This approach yields similar quantum advantages as in the case of solving PDEs \cite{Akshay2024Optimization}. That is when implementing the power method on tensor networks, the ranks can grow significantly, while for a quantum computer, the complexity will not depend on the ranks $y$ during the multiplications.

\subsection{Machine Learning}

It is possible to consider arbitrary neural networks (convolutional neural networks, transformers, etc.) in the form of tensor networks~\cite{naumov2023tetra, laskaris2023comparison, abronin2024tqcompressor}.
The contraction of these tensor networks would be computationally challenging on a classical computer due to the presence of large bond dimensions~\cite{markov2008simulating}. 
However, by encoding these tensor networks on a quantum computer, the calculations could be performed efficiently regardless of the architecture's complexity, which is a promising avenue for quantum machine learning~\cite{melnikov2023quantum}.

\subsection{Quantum Chemistry}

The description of molecules involves an accurate simulation of electrons and nuclei. The most common representation is the nonrelativistic molecular Hamiltonian
\begin{equation}
    \hat H = \sum_N \hat T_N + \sum_e \hat T_e + \sum_{N,M} \frac{Z_N\cdot Z_{M}}{\hat r_{NM}} -\sum_{N, e}\frac{Z_N}{\hat r_{Ne}}+\sum_{e,f}\frac{1}{\hat r_{ef}}.
\end{equation}
Here, $N, M$ denote nuclei, $e, f$ denote electrons, $\hat T$ denotes the kinetic energy, $\hat r$ is the distance of two particles, and $Z$ is the nuclear charge. In this formulation, both the electrons and the nuclei are described as quantum particles and they interact via the Coulomb force. The adiabatic or Born-Oppenheimer approximation splits this problem into weakly coupled electron and nuclei sub-problems. While certain problems require including relativistic effects or going beyond the Born-Oppenheimer approximation, a large class of quantum chemical problems are solved this way.

After invoking the Born-Oppenheimer approximation, the problem is reduced to the so-called electronic structure problem. In the second quantization formulation, the corresponding Hamiltonian reads
\begin{equation}
    \hat H_{el} = \sum_{p,q} h_{pq}\cdot \hat a_p \hat a_q^\dagger + \sum_{p,q,r,s} h_{pqrs} \cdot \hat a_p \hat a_q \hat a_r^\dagger \hat a_s^\dagger.
\end{equation}
Here, $p, q, r, s$ are atomic orbital indices, $\hat a, \hat a^\dagger$ are fermionic annihilation and creation operators, respectively, $h_{pq}$ entails the kinetic energy of the electrons plus the nuclear-electron attraction for fixed nuclear positions, and $h_{pqrs}$ encodes electron-electron repulsion. The Hamiltonian elements stem from the corresponding integrals over atomic orbitals and are evaluated using recursion formulae~\cite{helgaker2013molecular}. The brute force approach to solving the problem scales exponentially with the number of electrons and orbitals. DMRG has been a powerful tool in treating the electronic structure problem~\cite{chan2011density,chan2016matrix}. In this approach, the electronic state is described as a TT, and self-consistent field schemes are used to solve for ground- and excited states.

The quantum tensor network formulation of this problem facilitates simulating strong entanglement cases on a quantum computer: first, a DMRG computation is performed to obtain a low-rank approximation to the system of interest. This serves as the initial state. Subsequently, the MPO is represented using a unitary MPO and mapped to a quantum circuit using the scheme described in the present work.

The electron-electron correlation determines the required rank in DMRG methods and varies widely between molecules, their geometries, and considered states. For certain instances of the problem, the tensor train description exceeds classical resources and sufficiently reliable quantum computers can provide an advantage in solving these systems. While ground state chemistry of standard organic molecules~\cite{lee2023evaluating} is often described sufficiently accurately using polynomially scaling methods such as coupled cluster and/or perturbation theory, certain systems are beyond the capabilities of today's classical computers. Potential candidate systems include metal complexes, e.g. chromium dimer, active sites of proteins with metal atoms in their center, and systems showing nonadiabatic interactions in which multiple electronic states are interacting. The effectiveness of quantum computers still depends on the robustness of the quantum computers and certain TN structures might provide more accurate solutions at lower ranks.

\section{Conclusion}\label{sec:conclusion}

This article introduces a novel numerical method for encoding MPOs into quantum circuits with a depth that depends linearly on the number of qubits.
The devised algorithm offers a compelling solution to the crucial challenge of efficiently encoding matrices into quantum systems. It enables the effective encoding of a wide variety of matrices, not strictly confined to unitary matrices, that can be represented in MPO form. Such matrices are prevalent in diverse practical applications, ranging from solving partial differential equations to optimization.

To test the algorithm's applicability, we addressed the encoding of matrices like the Laplace operator, a diagonal matrix, the multi-controlled Toffoli gate, and the evolution operator of the Ising model Hamiltonian. Our findings illustrate the high-precision encoding of these matrices onto quantum circuits, demonstrating the algorithm's efficiency for matrices that can act on up to 50 qubits. This demonstrates the algorithm's versatility and potential to enhance quantum computational capabilities in diverse domains. Notably, the proposed circuits exhibit a linear scaling of gate count with qubit number, establishing the efficiency of our matrix-encoding algorithm in comparison to existing multi-qubit gate decomposition methods, which have exponential scaling~\cite{TheorLimQGdec, k-qubits_gate, QuantumShannonDecomposition}.

An effective solution to the problem of encoding matrices into a quantum circuit using TTs allowed us to introduce the concept of hybrid Tensor Quantum Programming. This strategy integrates classical and quantum computing to solve numerical problems framed within tensor networks. Starting with a classical phase, computations are performed classically, while the TT-ranks are small, benefiting from the efficiency of classical implementation. When the ranks increase, indicating potential quantum benefits, the quantum phase is initiated and the intermediate solution is transferred to a quantum computer. During this phase, the rank growth is unlimited, and once the ranks decrease again, the process reverts to the classical phase. This hybrid approach outperforms purely classical methods by efficiently handling large-rank problems, utilizing classical computation for small ranks and quantum computation for large ranks. Consequently, it optimizes computational resources for addressing problems with varying degrees of entanglement and correlation.

It is important to note that while our MPO-encoding algorithm demonstrates good applicability, particularly in scenarios featuring low-rank MPOs, it does not completely solve the problem of accurately encoding arbitrary matrices into quantum circuits. In the general case, an arbitrary matrix has full rank in the MPO format, which can significantly complicate the encoding scheme. However, our algorithm can be successfully used for encoding a low-rank MPO approximation of the original matrix obtained by cutting the rank with the introduction of an additional error. 

The utilization of Riemannian optimization allows us to achieve high levels of accuracy in our MPO-encoding algorithm. However, such optimization techniques do not inherently guarantee the absolute accuracy of the obtained solution. Using block-encoding for the MPO~\cite{nibbi2023block} provides completely accurate encoding but requires the incorporation of an additional auxiliary qubit for each MPO core. Consequently, the number of auxiliary qubits increases in proportion to the system qubits, which limits the applicability of this method, particularly for encoding large matrices. Riemannian optimization allows for more flexible MPO-encoding, which is imprecise but requires fewer auxiliary qubits. Thus, a promising path for future development may be exploring the trade-off between the encoding accuracy, the auxiliary qubit number, and the probability of successful implementation. This is achievable by improving the optimization strategy in our algorithm, combining it with block-encoding, or developing completely new encoding techniques.

The success probability in the MPO encoding algorithm is determined not only by the intrinsic properties of the encoded matrix itself and the encoding accuracy but also by the initial state vector on which the matrix acts. For specific initial states, the success probability can significantly exceed the average success probability. The highest success probability is achieved when the initial state aligns with the encoded matrix's right singular vector with the maximum singular value. Such initial states prove important not only for maximizing success probability but also from a practical standpoint, since finding the largest singular value (or eigenvalue) states is a prevalent challenge. 
In this case, following the Tensor Quantum Programming scheme, the initial state is a classically precomputed approximate solution, which is close to the largest singular-value state.
This approach can greatly increase the likelihood of successfully implementing the quantum part, finding utility, for instance, in approaches like the optimization power method \cite{chertkov2022optimization_power}. 

This work opens up directions for further optimal development of quantum algorithms and computations. The driving idea is to find the drawbacks and difficulties in tensor network algorithms and try to eliminate them with the help of quantum computers, drawing on the substantial connection between both approaches. The primary example is a significant increase in the ranks inherent in numerical schemes, which can be minimized by the ability of a quantum computer to operate on states with arbitrary entanglement (ranks).

Our concept of Tensor Quantum Programming is universal and can be generalized to different tensor network architectures \cite{khrulkov2017expressive, reyes2021multi}. It can be tailored to particular problems and available quantum device architectures with specified qubit connectivity. This adaptability allows the Tensor Quantum Programming scheme to find potential applications in a range of fields coupled to tensor networks, for instance, in machine learning. Further research will be required to explore these avenues.

\begin{acknowledgments}
We thank Alexey Melnikov, Dmitry Morozov, Akshay Vishwanathan, Vyacheslav Kuzmin, and Valerii Vinokur for useful discussions.
\end{acknowledgments}

\clearpage

\bibliographystyle{unsrt}
\bibliography{bib}

\end{document}